\documentclass[a4paper,psamsfonts,twocolumn]{article}
\usepackage{amsmath}
\usepackage{amssymb}
\def\ifnextchar#1#2#3{\let\tmpnce=#1%
    \def\tmpnca{#2}\def\tmpncb{#3}\futurelet\tmpncc\ifnch}%
  \def\ifnch{\ifx\tmpncc\tmpnce\let\tmpncd=\tmpnca%
	\else\let\tmpncd=\tmpncb\fi\tmpncd}
\def\d{\operatorname{d}\!}

\def\monthname{\ifcase\month\or Jan\or Feb\or March\or Apr\or %
    May\or June\or July\or Aug\or Sept\or Oct\or Nov\or Dec\fi}
\def\norm#1{\left\|#1\right\|}
      \def\normp#1_#2{\norm{#1}_{#2}}

\let\R\Real

\input epsf.tex  
\def\caption#1{\hfill \\
  \hbox{}\hfil{\footnotesize #1}\hfil\hbox{}}
\def\ifnextchar#1#2#3{\let\tmpnce=#1%
    \def\tmpnca{#2}\def\tmpncb{#3}\futurelet\tmpncc\ifnch}%
\def\ifnch{\ifx\tmpncc\tmpnce\let\tmpncd=\tmpnca%
    \else\let\tmpncd=\tmpncb\fi\tmpncd}
\def\tpeinture #1 by #2 (#3){
  \vtop to #2{
    \hrule width #1 height 0pt depth 0pt
    \vfill
    \epsfysize=#2 \epsfbox{#3}
    }
  }
\def\xpeinture #1 by #2 (#3){
  \hbox{$\vcenter to #2{
    \hrule width #1 height 0pt depth 0pt
    \vfill
    \epsfxsize=#1 \epsfbox{#3}
    }$}
  }
\def\ypeinture #1 by #2 (#3){
  \hbox{$\vcenter to #2{
    \hrule width #1 height 0pt depth 0pt
    \vfill
    \epsfysize=#2 \epsfbox{#3}
    }$}
  }
\def\peinture{\ypeinture}
\def\bpeinture #1 by #2 (#3){
  \vbox to #2{
    \hrule width #1 height 0pt depth 0pt
    \vfill
    \epsfysize=#2 \epsfbox{#3}
    }
  }
\def\Scaledpiu[#1] #2 by #3 (#4){{ %
   \dimen0=#2 \dimen1=#3 %
   \if#1t
       \tpeinture \dimen0 by \dimen1 (#4)%
   \else\if#1b
        \bpeinture \dimen0 by \dimen1 (#4)%
        \else\if#1x
             \xpeinture \dimen0 by \dimen1 (#4)%
             \else
             \ypeinture \dimen0 by \dimen1 (#4)%
            \fi\fi
    \fi}}
\def\Scaledpiv #1 by #2 (#3){{%
   \dimen0=#1 \dimen1=#2%
   \peinture \dimen0 by \dimen1 (#3) %
   }}
\def\scaledpicture{\ifnextchar[{\Scaledpiu}{\Scaledpiv}}
\def\centredpicture #1 by #2 (#3){
   \par\centerline{\hbox{
   \scaledpicture #1 by #2 (#3)}
   }}

   \input epsf.tex
\title{\large \bf Physical Parameters for Biconcave Shape Vesicles}

\author{\normalsize Thomas Kwok-keung Au and Tom Yau-heng Wan \\
{\small{\em Department of Mathematics, The Chinese University of Hong Kong, Shatin, Hong Kong}} \\ 
\parbox{16cm}{\small
\vskip 2ex {\bf Abstract.}
The Helfrich's shape equation of axisymmetric vesciles is studied.  A sufficient condition on the physical parameters and some geometric properties are discovered for the formation of biconcave shape vesicles.}
}
\date{}

\def\HelF{{\mathcal F}}
\def\tlambda{\tilde{\lambda}}
\def\tp{\tilde{p}}

\begin{document}
\maketitle

\markright{Physical Review Letters, \today}
\setlength{\baselineskip}{14pt}

The study of the biconcave-discoid shape of a red blood cell has been a continued interest in the last two decades.  Theoretically, the shape of the cell is to minimize a certain bending energy.  Like minimal surfaces and surfaces with constant mean curvature in geometry, the expected biconcave shape surface is a critical one for the bending energy functional.  The early proposal for the bending energy by Canham \cite{Canham} is purely geometric, which mainly involves the Willmore functional, \cite[ch.~7]{Willmore}.
It is well known among differential geometers that the unique minimum of the Willmore functional for topologically spherical vesicles is the round sphere, which differs from blood cells observed experimentally.  Thus, Willmore functional or Canham's idea is not a good model for the shape of red blood cells.  This is also observed by physicists \cite{Helfrich1976}.

It is known that the shape of blood cells and other biological membranes is closely related to the formation of lipid bilayer vesicle in aqueous medium.  Based on this and the elasticity of lipid bilayer, Helfrich proposed a modified bending energy, \cite{Helfrich1973}.  For a closed surface $\Sigma$, this bending energy is a combination of geometric quantities and physical parameters.  The geometric quantities are the volume $\text{V}(\Sigma)$ enclosed by $\Sigma$, the area $\text{A}(\Sigma)$, the mean curvature $H$ and the Gaussian curvature $K$ of $\Sigma$.  According to the Gauss-Bonnet Theorem, the integral of $K$ is a topological constant.  Within a certain topological class of $\Sigma$, Helfrich's bending energy can be reduced to, which we call Helfrich functional,
$$
\HelF(\Sigma) = \oint_\Sigma (2H+c_0)^2 \d A + \tlambda \text{A}(\Sigma) + \tp\text{V}(\Sigma)
$$
where $c_0$, $\tlambda=2\lambda/k_c$, and $\tp = 2p/k_c$ are the physical constant parameters of the functional interpreted as: $k_c$ is the bending rigidity, $c_0$ the spontaneous curvature, $\lambda$ the tensile stress, and $p=p_o-p_i$ the osmotic pressure difference between the outer ($p_o$) and inner ($p_i$) media.  Here, we have taken a sign convention such that $H$ is negative for spheres and Helfrich's numercial simulation produces a biconcave shape when $c_0$ is positive.

The general minimizer of $\HelF$ is not easy to find.  In the past, much effort has been spent on studying axisymmetric solution to its variational equation.  This equation is usually referred to as {\sl shape equation of axisymmetric vesicles}.  There are numerous works on axisymmetric stationary vesicles, \cite{Helfrich1976,Luke,Seifert,Mutz-Bensimon,OuYang-Helfrich1989,OuYang,Naito-Okuda-OuYang1,Naito-Okuda-OuYang2}.  It is still unknown to scientists what conditions on the physical parameters will guarantee a solution to the shape equation corresponding to a biconcave surface.

In this letter, we provide a sufficent condition under which the Helfrich shape equation of axisymmetric vesicles has solutions of biconcave shape.  Besides, we also report two necessary geometric conditions for the lipid bilayer vesicles, one for those with a reflection symmetry and another for those of biconcave shape.
These conditions on the parameters are very mild but easy to verify.  In particular, the case that $c_0>0$, $\tlambda >0$, and $\tp >0$ is sufficient to ensure the existence of biconcave  solution.  The discussion on how the combination of the parameters $c_0$, $\tlambda$, and $\tp$ affects the existence of biconcave solution will be presented in another review paper in detail.

The sufficient condition for the formation of biconcave vesicle may be formulated in terms of a cubic polynomial,
$$
Q(t) = t^3 + 2c_0 t^2 + (c_0^2 + \tlambda)t - \frac{\tp}{2}.
$$
We proved that {\bf if all roots of $Q(t)$ are positive, then one can always find axisymmetric biconcave vesicles which are the stationary surface for the Helfrich functional}.  A pictorial summary is given at the end of this letter (p.~\pageref{summarytable}).

For the necessary conditions, we observe that if $c_0>0$, the central curvature of an axisymmetric stationary vesicle of biconcave shape must be smaller than the first positive root of $Q(t)$, more precisely, see~(\ref{upperw0}).  
Furthermore, with the reflection symmetry orthogonal to the rotation symmetry, the Gaussian curvature along the ``equator'' (the intersection circle of the axisymmetric vesicle with the plane of reflection) is explicitly given in terms of the radius $r_\infty$ of the ``equator" and the polynomial $Q(t)$, see~(\ref{equatorK}).

\subsubsection*{Equation and Requirements}
For an axisymmetric surface $\Sigma$, we take the rotational axis to be $z$-axis and the plane of reflection $xy$-plane, then a biconcave shape is obtained by revolving and reflecting a curve, given by a function $z(r)\geq 0$ defined for $r$ in some interval $[0,r_\infty]$, where $r_\infty$ is the radius of the ``equator''.  For a biconcave axisymmetric surface, we expect $z(r)$ and its derivative $w=z'$ to have the following graphs.
\begin{center}
\mbox{\xpeinture 3.2cm by 2.1cm (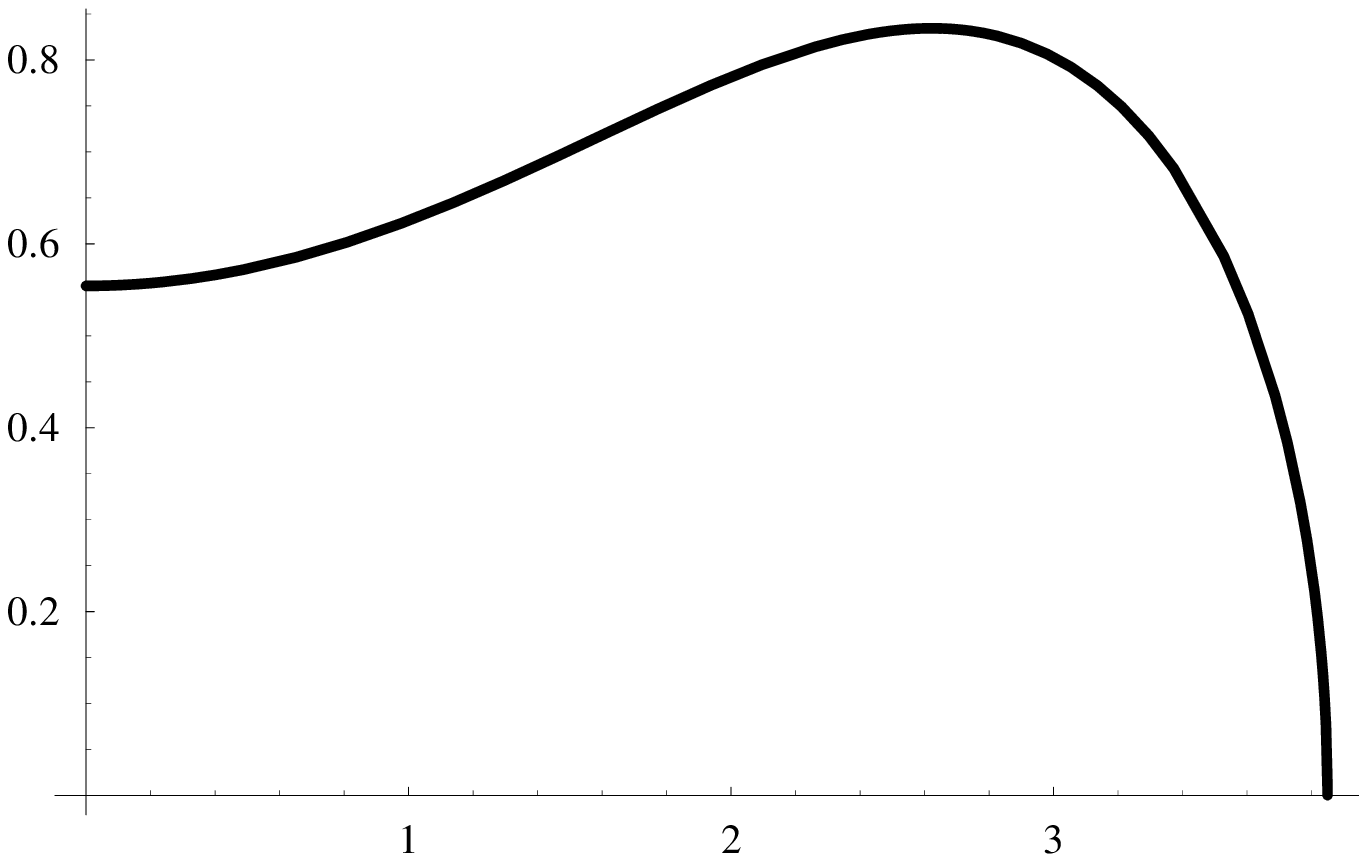)}
\hfil
\mbox{\xpeinture 3.2cm by 2.1cm (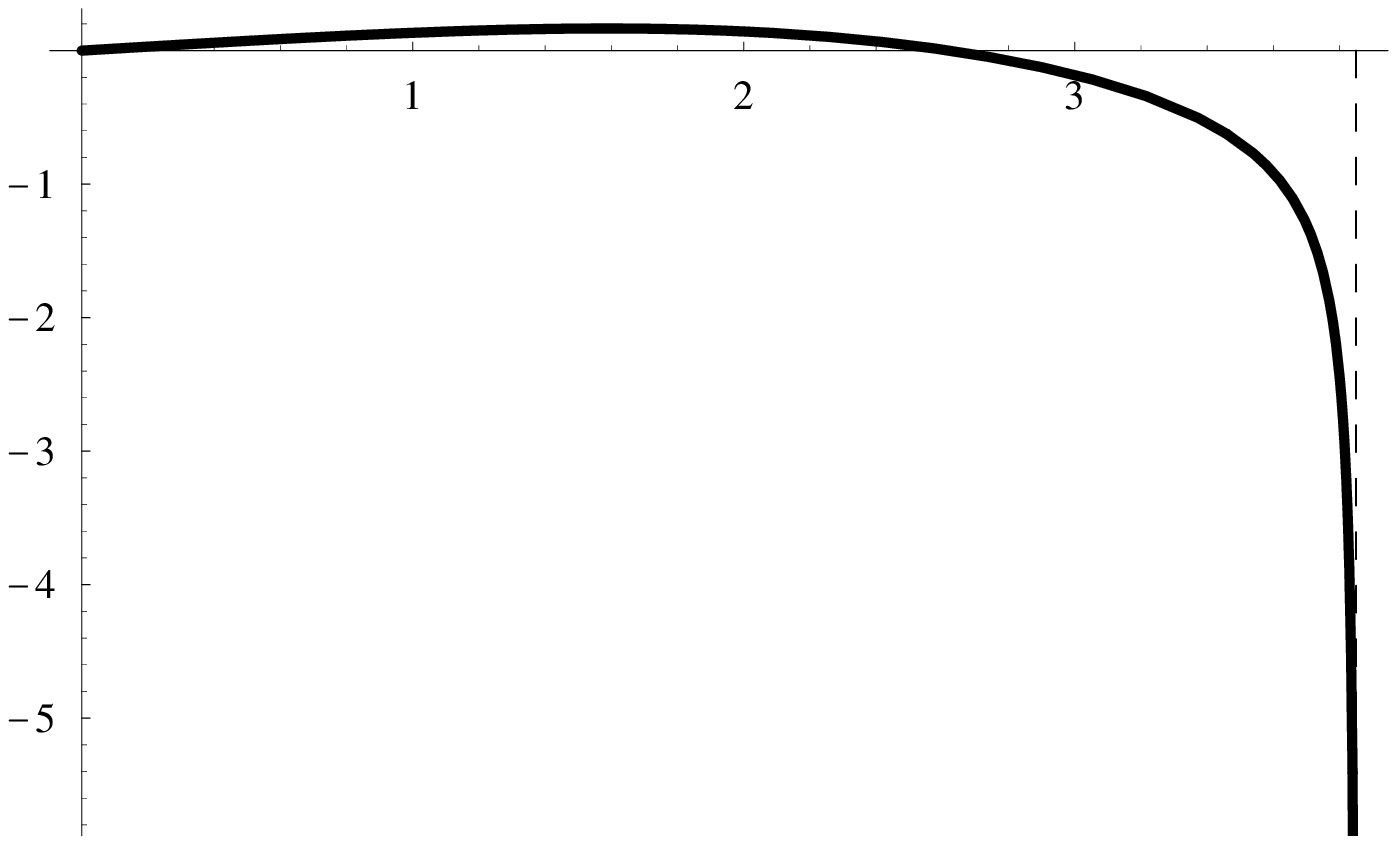)}
\end{center}

Usually, the Helfrich shape equation of axisymmetrc vesicles is written in term of the angle $\psi$ between the surface tangent and the plane perpendicular to rotational axis 
\cite{Helfrich1976,Julicher-Seifert,Zheng-Liu}.  For our discussion, it is convenient to rewrite it in terms of the derivative of the graph $w(r)=z'(r)=\tan \psi$. Then the Helfrich shape equation of axisymmetrc vesicles becomes an equation for $w$, which is
\begin{equation} \label{VEqn1}
\begin{aligned}
\frac{2r{w''}}{(1+w^2)^{5/2}} &=  \frac{5rw{w'}^2}{(1+w^2)^{7/2}} \\
&\hbox{}\hspace*{-1.5em} - \frac{2{w'}}{(1+w^2)^{5/2}}  + \frac{2w+w^3}{r(1+w^2)^{3/2}} \\
&\hbox{}\hspace*{-1em} + \frac{2c_0 w^2}{1+w^2} + \frac{(c_0^2+\tlambda)r w}{(1+w^2)^{1/2}} - \frac{\tp r^2}{2}.
\end{aligned}
\end{equation}
The biconcave shape and symmetries are then translated into suitable initial and boundary conditions, such as,
$w(0)=0$; $w'(0)={w_0'} > 0$; $w(r)\to -\infty$ as $r\to r_\infty$; an integral condition that $-\infty \ne \displaystyle \int_0^{r_\infty} w\d r < 0$; and $w$ is required to have a unique local maximum and no other critical points.

The lower order terms can be given by polynomials $Q(t)$ or its quadratic part $R(t)=Q(t)-t^3$.
After multiplying with $r {w'}$, the equation~(\ref{VEqn1}) can be written as follow,
\begin{gather}
\begin{aligned}
\left[ \frac{r^2 {w'}^2}{(1+w^2)^{5/2}} \right]' &= \left[ \frac{w^2}{(1+w^2)^{1/2}} \right]' \\
&\qquad {} + r^3 {w'} R(\kappa(r));
\end{aligned} \label{VEqn2}\tag{\ref{VEqn1}a} \\
\begin{aligned}
\left[ \frac{r^2 {w'}^2}{(1+w^2)^{5/2}} \right]' &= \left[ \frac{-2}{\sqrt{1+w^2}} \right]' \\
&\qquad {} + r^3 {w'} Q(\kappa(r)). 
\end{aligned} \label{VEqn3}\tag{\ref{VEqn1}b}
\end{gather}
where $\kappa(r) = \dfrac{w}{r\sqrt{1+w^2}}$  is the principal curvature of $\Sigma$ in the meridinal direction.  Indeed,
$$
\kappa(r) \to {w_0'}, \qquad \text{as $r\to 0$,}
$$
so the initial choice ${w_0'}$ is the central meridinal curvature or ${w_0'}^2$ the Gaussian curvature.
Note that if $c_0 = \tlambda = \tp = 0$, equation (\ref{VEqn2}) can be integrated directly to a solution $z=z(r)$ which is a circular arc with radius $1/{w_0'}$.  This is exactly the case when the meridinal curvature is a constant.

\subsubsection*{Necessary Conditions}
The lower order terms in~(\ref{VEqn1}) capture the changes of the meridinal curvature.
Since $Q(0) = R(0) = -\tp/2 < 0$, if ${w_0'}$ is less than the smallest positive root of $Q(t)$, then both $Q({w_0'}) < 0$ and $R({w_0'}) < 0$.  Our analysis shows that these negativity conditions are essential for the meridinal curvature $\kappa(r)$ to change sign.  Furthermore, if the physical parameters constitute a polynomial $Q$ with only positive roots, a solution $w(r)$ with $w_0'$ small enough always satisfies all of our requirements.  The threshold value of $w_0'$ can be calculated numerically if exact values of the parameters are given.

The meridinal curvature $\kappa(r)$ of the vesicle $\Sigma$ is the most basic quantity in our analysis.  From equation (\ref{VEqn3}), if $c_0>0$ and the initial curvature at the center $w_0'$ is large, in the sense that $R(w_0') > 0$, then $\kappa(r)$ increases and a biconcave shape cannot be formed.  So a necessary condition for formation of axisymmetric solution of biconcave shape in the case that $c_0>0$ is
\begin{equation}\label{upperw0}
R(w_0')=2c_0{w_0'}^2+ (c_0^2 + \tlambda)w_0' - \frac{\tp}{2}<0.
\end{equation}

The following pictures show the behaviors of $\kappa(r)$ for two different values of $w_0'$ with $c_0=1$, $\tlambda=0.25$, and $\tp=1$.  For these parameter values, $R(0.277124)=0$.
\begin{center}
\parbox{3.2cm}{\ypeinture 3.2cm by 2.1cm (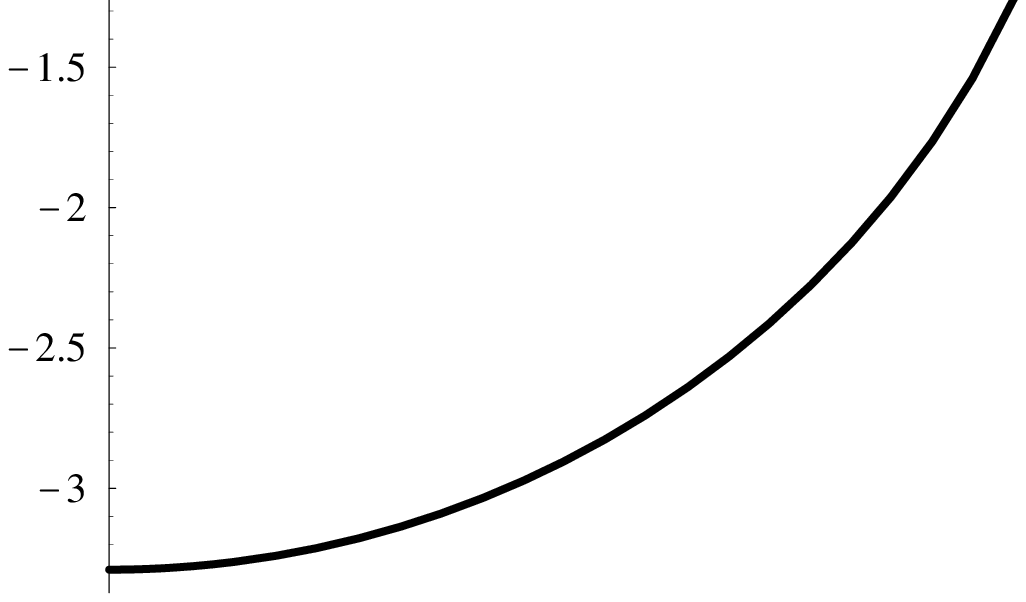) 
\caption{$R(w_0')>0$}
} \hfil
\parbox{3.2cm}{\ypeinture 3.2cm by 3cm (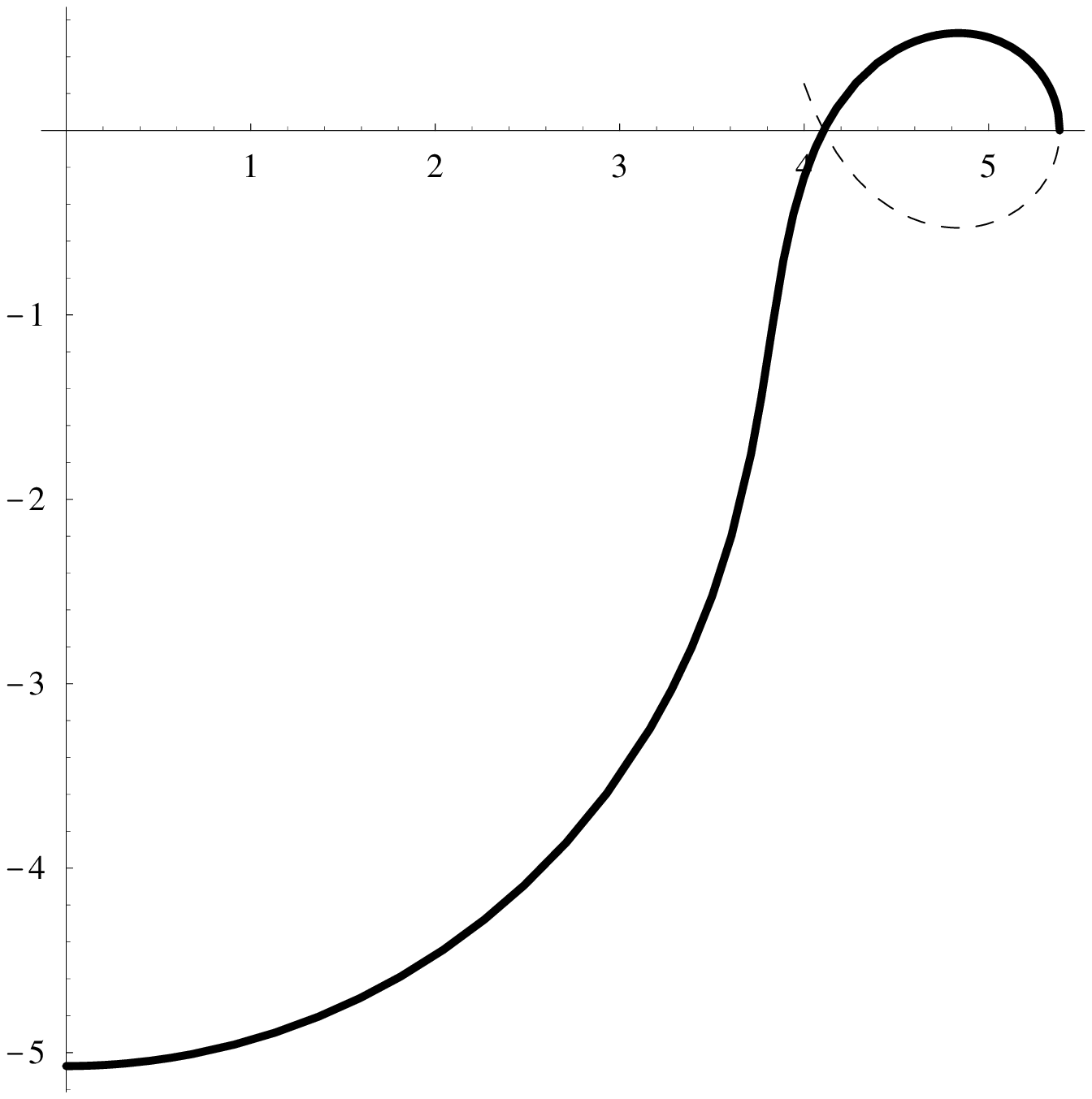) 
\caption{$R(w_0')<0$}
}
\end{center}
The one on the left is taken with $w_0'=0.278$ so the necessary condition is not satisfied. In this case, $\kappa$ blows-up at finite radial distance and the solution only leads to a convex surface.  The picture on the right is taken with $w_0'=0.276$.  In this case, the necessary condition is satisfied, so $\kappa$ decreases and changes sign.  But it only decays gradually and the integral area condition is not satisfied.  As a result, the surface and its reflection intersects in $\R^3$.

There is a more subtle boundary condition imposed by the reflection symmetry.  One needs to verify that $z(r)$ and $-z(r)$ together form a stationary solution with respect to any variation near the plane of reflection.  For the situation that we discuss, it turns out that this is true.  Moreover, we obtain a relation between the radius $r_\infty$ of the ``equator" and the Gaussian curvature $K(r_\infty)$ of the vesicle at any point on the ``equator",
\begin{equation}\label{equatorK}
K(r_\infty)^2 = \frac{-1}{r_\infty} Q\left(\frac{-1}{r_\infty}\right).
\end{equation}
This relation is easily verified for the case that $c_0=\tlambda=\tp=0$, in which, $Q(t)=t^3$ and $K(r_\infty)^2=\dfrac{1}{r_\infty^4}$. This is compatible with the fact that the solution surface is a round sphere.

\subsubsection*{Sufficient condition}
To have a solution meeting all the requirements, we need to impose a stronger condition that $w_0'$ to be small enough, at least smaller than the positive roots of $Q$.
This smallness assumption on $w_0'$ implies that not only $R(w_0')<0$ but also $Q(w_0')<0$.  Both $\kappa(r)$ and $w(r)$ start initially positive and become negative when $r$ reaches a certain value $r_0$ (depending on the value of $w_0'$, actually, comparable to ${w_0'}^{1/2}$ when $w_0'$ is small).
In fact, we discover that, for any point of inflection $r_c$ of the graph of $z(r)$ with $\kappa(r_c)>0$,
$$
\frac{16}{3\tp}\le \lim_{w_0'\to 0}\frac{r_c^2}{w_0'}\le\lim_{w_0'\to 0}\frac{r_0^2}{w_0'}\leq \frac{16}{\tp}.
$$
As a consequence of these inequalities, the graph of $z(r)$ has a unique point of inflection.
After $r$ reaches $r_0$, with the assumption that all roots of $Q(t)$ are positive, the lower order terms of equation~(\ref{VEqn3}) are dominated and the function $w(r)$ blows down monotonically to $-\infty$ in a finite radial distance.  By further estimating the area according to $w_0'$, the integral area condition can be verified whenever $w_0'$ is sufficiently small.  All these behaviors of the functions $z(r)$ and $w(r)$ with respect to $Q$ and the choice of $w_0'$ are illustrated in our summary on p.~\pageref{summarytable}.  The mathematical details will be presented in a forthcoming paper.

\subsubsection*{Discussions}
Finally, we would like to remark on other solutions which do not correspond to biconcave surfaces.
\begin{center}
\mbox{\xpeinture 3.2cm by 2.1cm (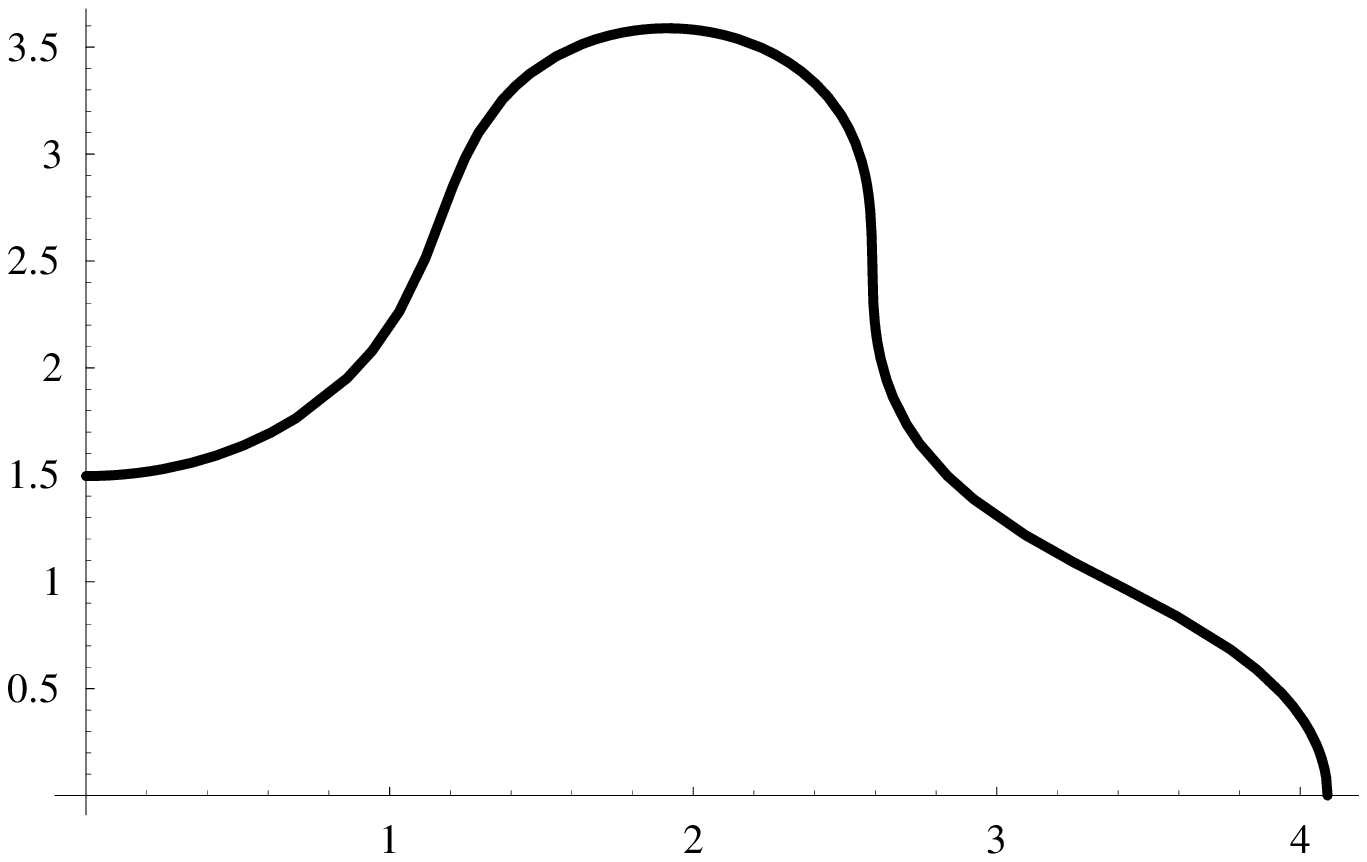)}
\hfil
\mbox{\xpeinture 3.2cm by 3.2cm (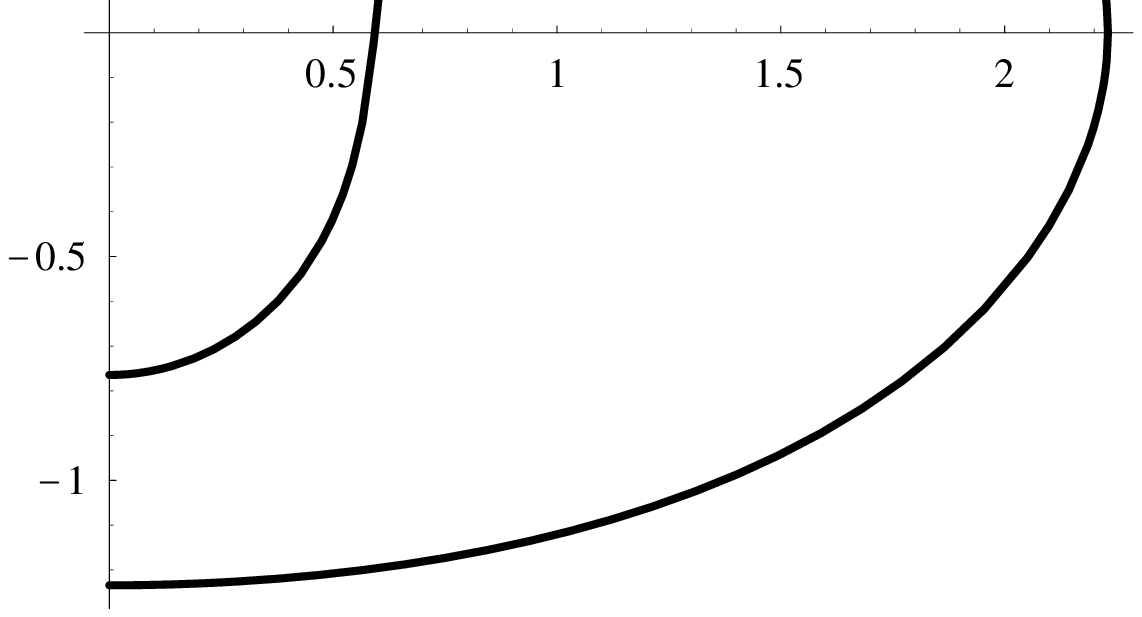)}
\end{center}
The one on the left is a solution for $c_0=1$, $\tlambda=-3$, $\tp=2$ with $w_0'=2$.  Note that $w_0'$ is greater than the positive roots of $R(t)$ and $Q(t)$.  Moreover, $Q(t)$ has negative roots and thus this is beyond the situation discussed above.  In this case, after $w=z'$ becomes negative, it may not decrease monotonically to negative infinity, which is why a second ``petal'' may form.

The other type is a rotational solution which is not symmetric under reflection in $xy$-plane \cite{Helfrich1976,Helfrich1977}.  The picture is produced with $c_0 = -1$, $\tlambda = -1$ and $\tp = 1$.
The ``upper'' curve satisfies our equation~(\ref{VEqn1}) but not the area condition.  The ``lower'' curve and the reflected ``upper'' curve are both solutions locally, so there is a bifurcation of solution at $r_\infty$.  The surface only closes up by an asymmetric one.  It should be remarked that, from our numerical study, not all solutions in the upper quadrant may be closed up by an asymmetric counterpart.  For example, if the upper part satisfies the area condition, the only possible lower part is the reflection symmetric one.  This suggests a possible uniqueness of biconcave solution.  However, at the moment, there is not yet a mathematical proof.

\smallskip

\noindent\label{summarytable}
\begin{tabular}{||ccc||}
\hline\hline
\multicolumn{3}{||c||}{\parbox{6.5cm}{
\begin{center}
Graph of $Q$ and positions of $w_0'$ \\
($c_0 = 1$, $\tlambda = 0.25$, $\tp = 1$) \\
\mbox{\xpeinture 6cm by 4cm (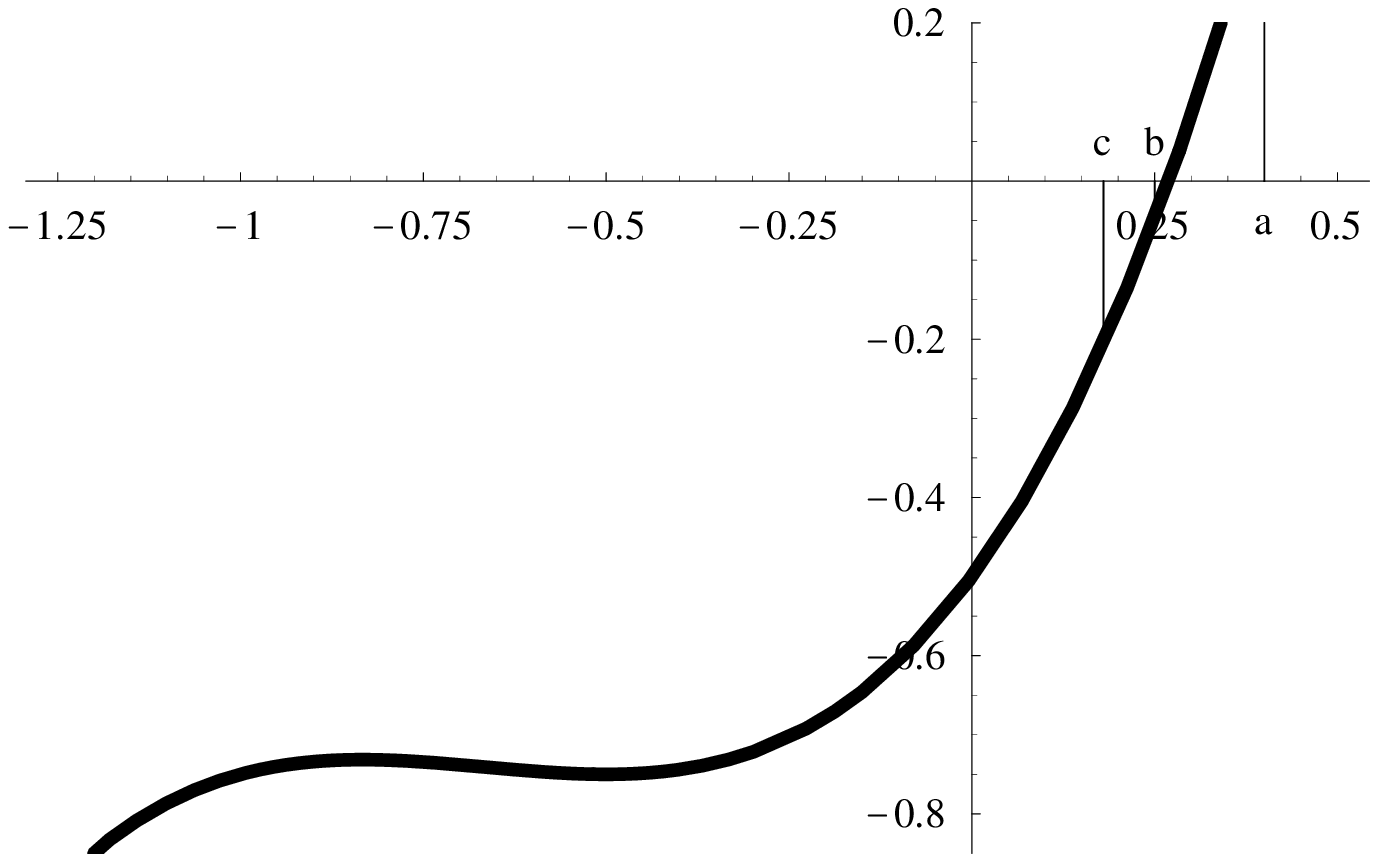)}
\end{center}
}} \\
\hline
\multicolumn{3}{||c||}{Graphs of solution $z$} \\
{\tiny (a) $w_0'=0.4$} 
& {\tiny (b) $w_0'=0.25$}
& {\tiny (c) $w_0'=0.18$}
\\
\parbox{2cm}{\xpeinture 2cm by 1.4cm (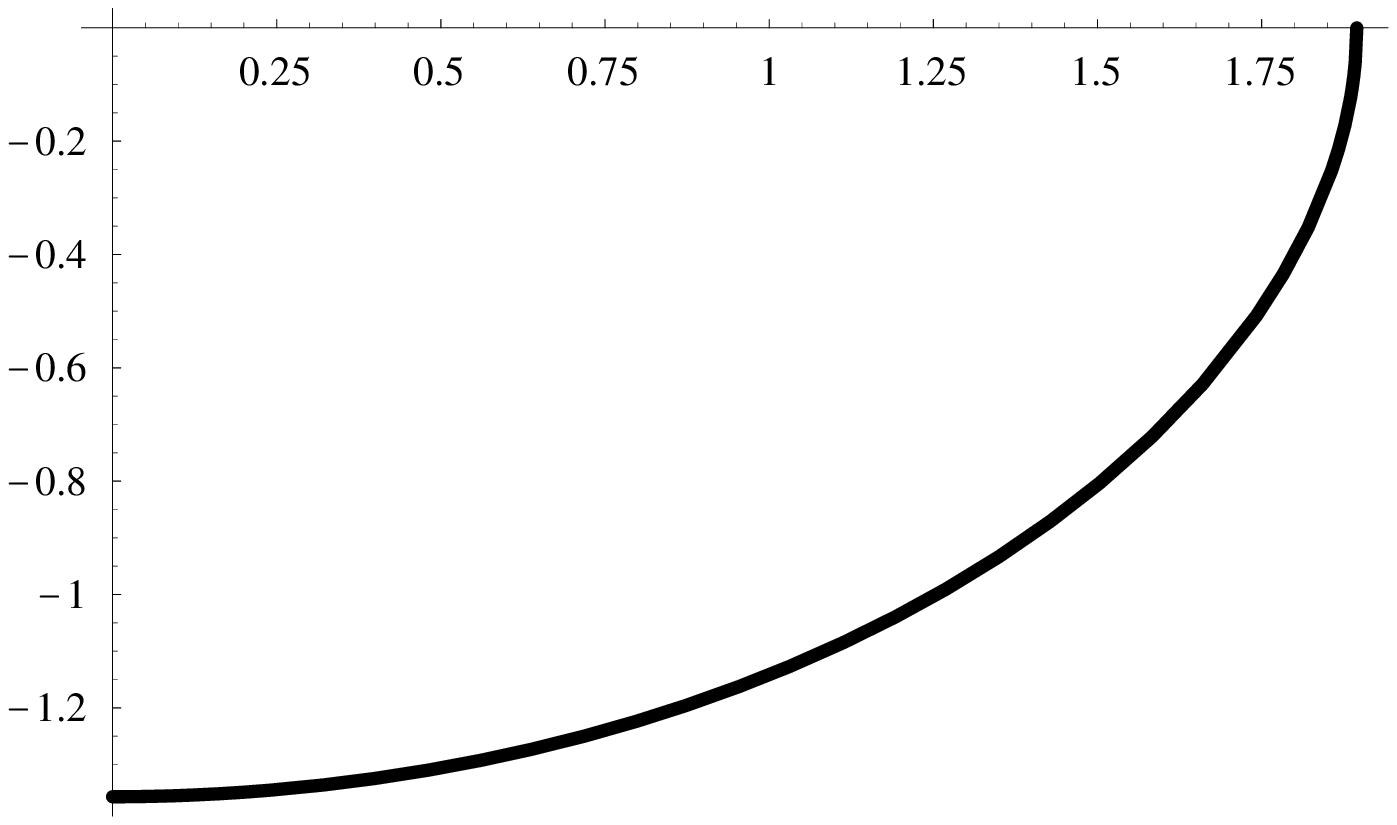)}
&\parbox{2cm}{\xpeinture 2cm by 1.4cm (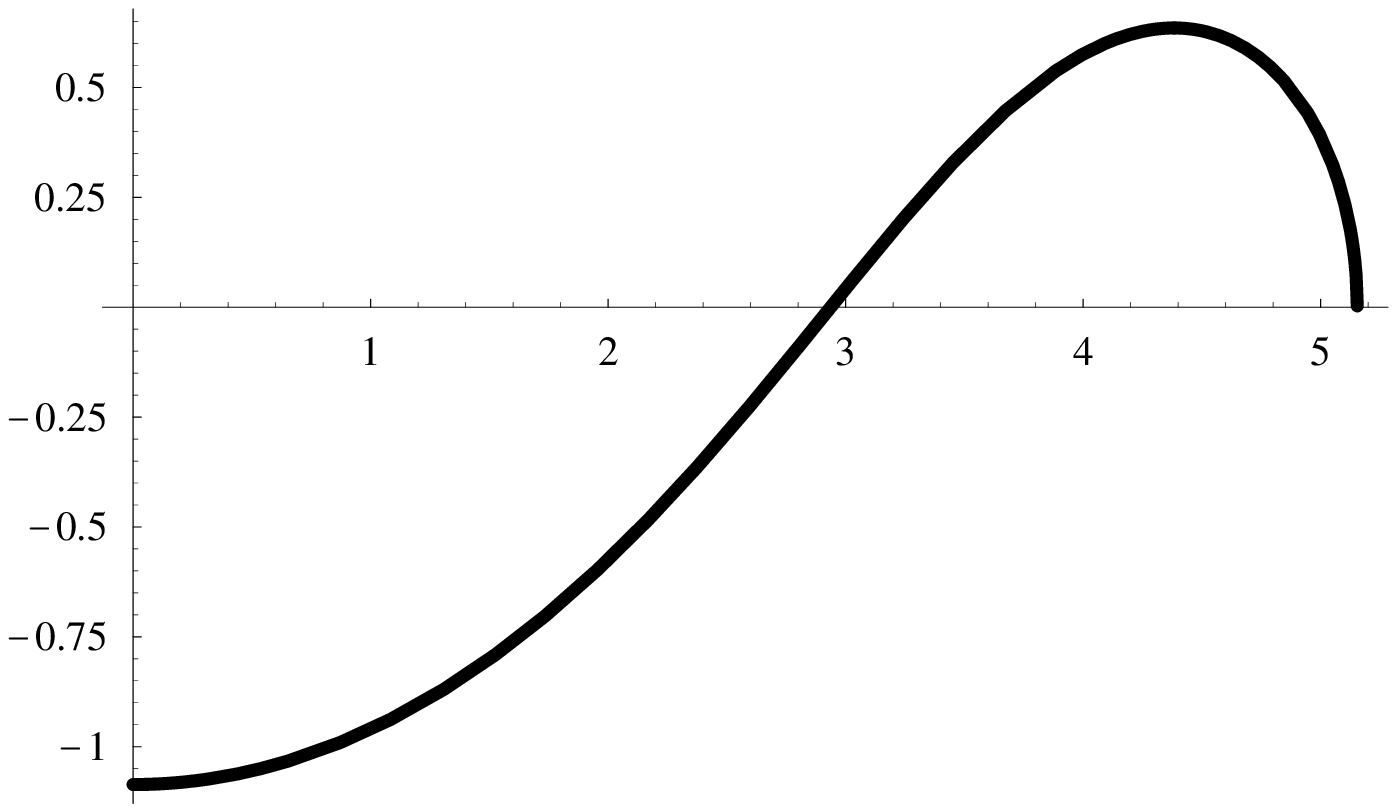)}
&\parbox{2cm}{\xpeinture 2cm by 1.4cm (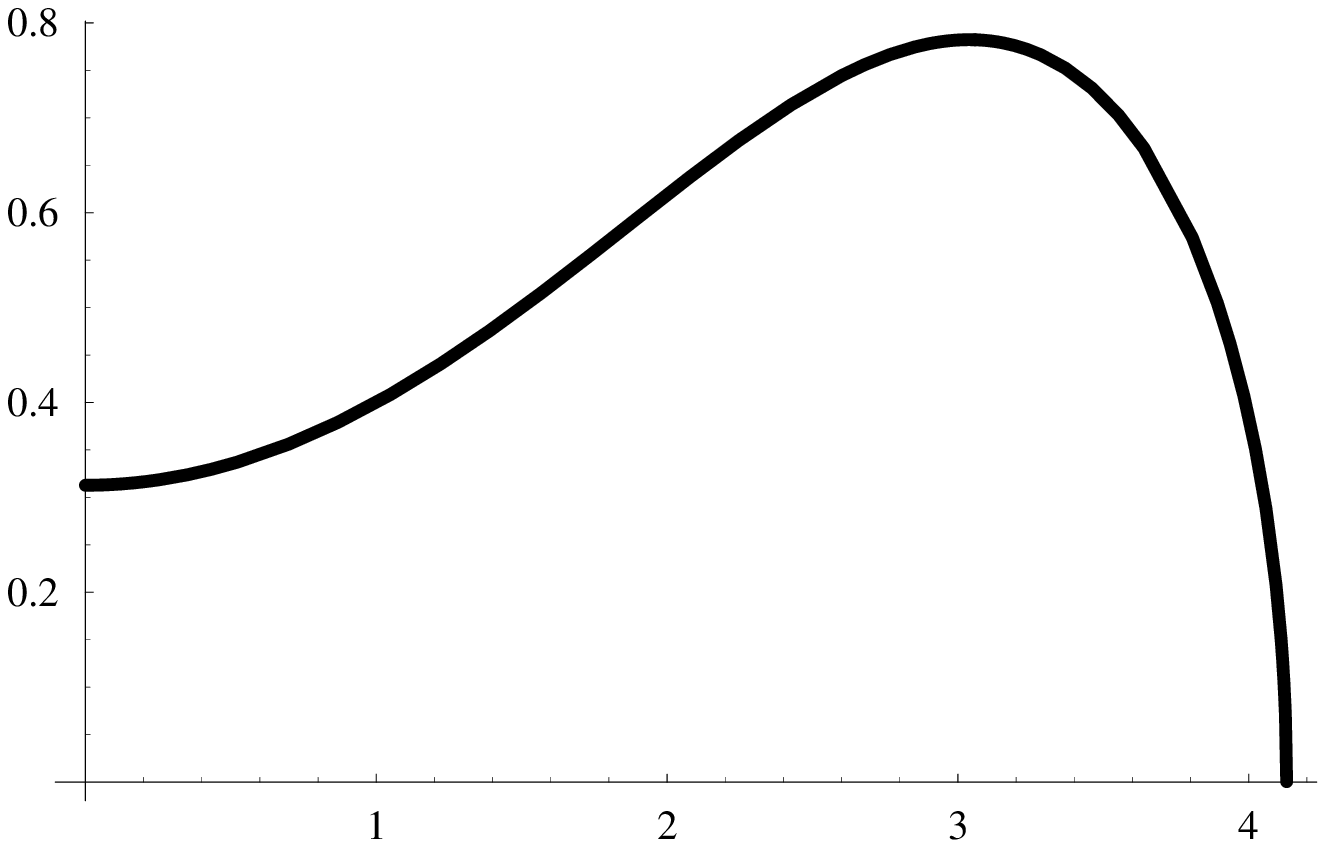)} \\
\hline\hline
\end{tabular}

\smallskip

\noindent
\begin{tabular}{||ccc||}
\hline\hline
\multicolumn{3}{||c||}{\parbox{6.5cm}{
\begin{center}
Graph of $Q$ and positions of $w_0'$ \\
($c_0 = 1$, $\tlambda = -0.3$, $\tp = 1$) \\
\mbox{\xpeinture 6cm by 4cm (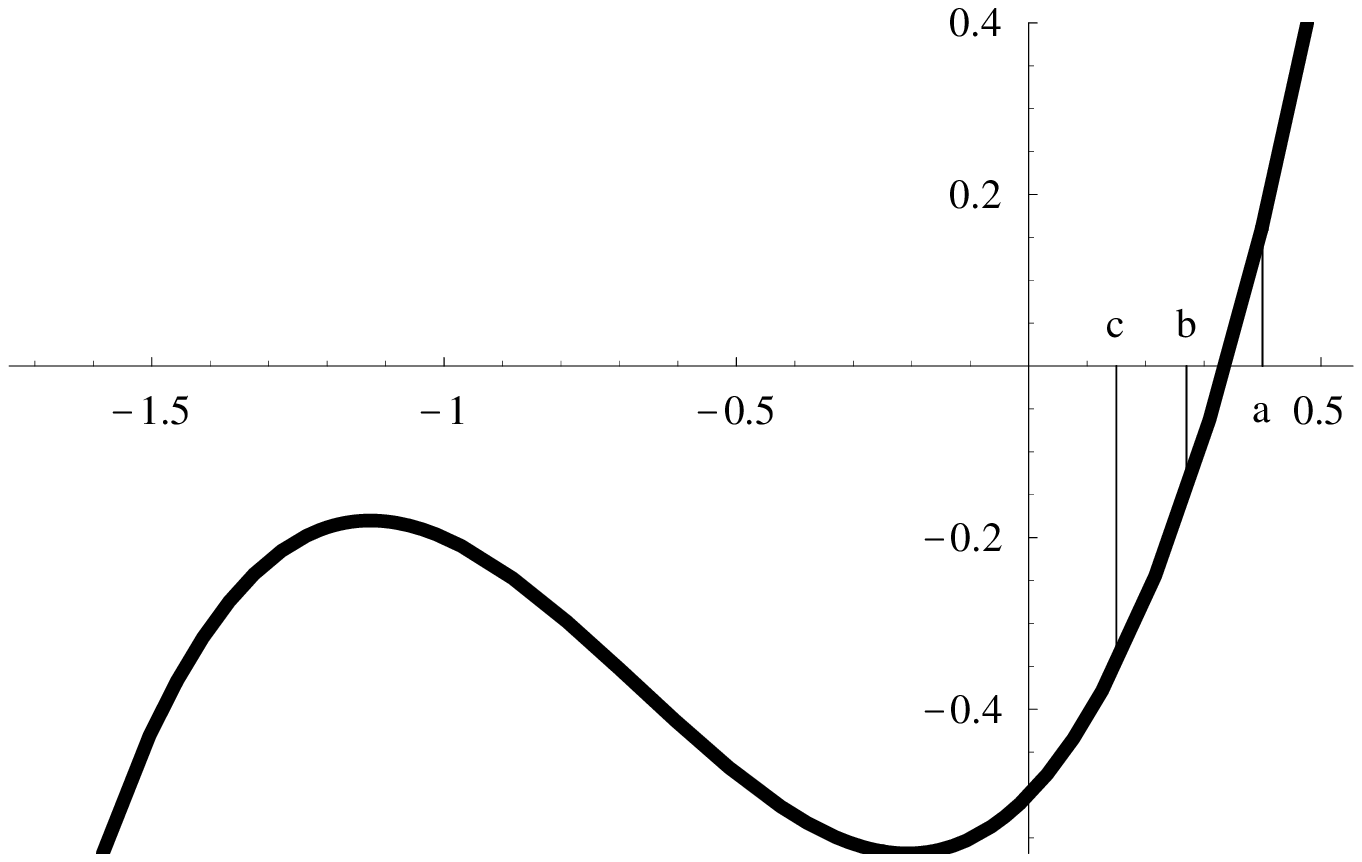)}
\end{center}
}} \\
\hline
\multicolumn{3}{||c||}{Graphs of solution $z$} \\
{\tiny (a) $w_0'=0.4$} 
& {\tiny (b) $w_0'=0.27$}
& {\tiny (c) $w_0'=0.15$}
\\
\parbox{2cm}{\xpeinture 2cm by 1.4cm (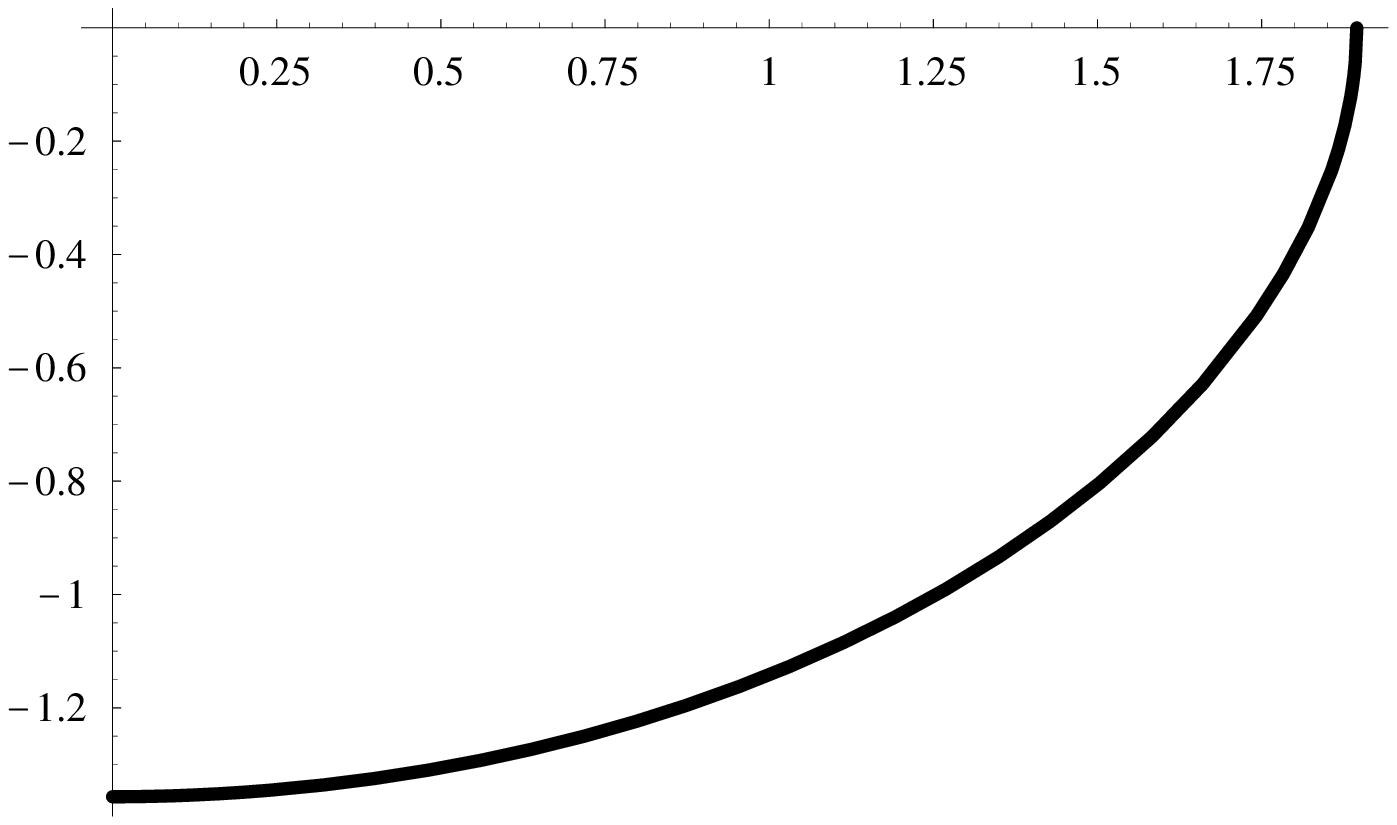)}
&\parbox{2cm}{\xpeinture 2cm by 1.4cm (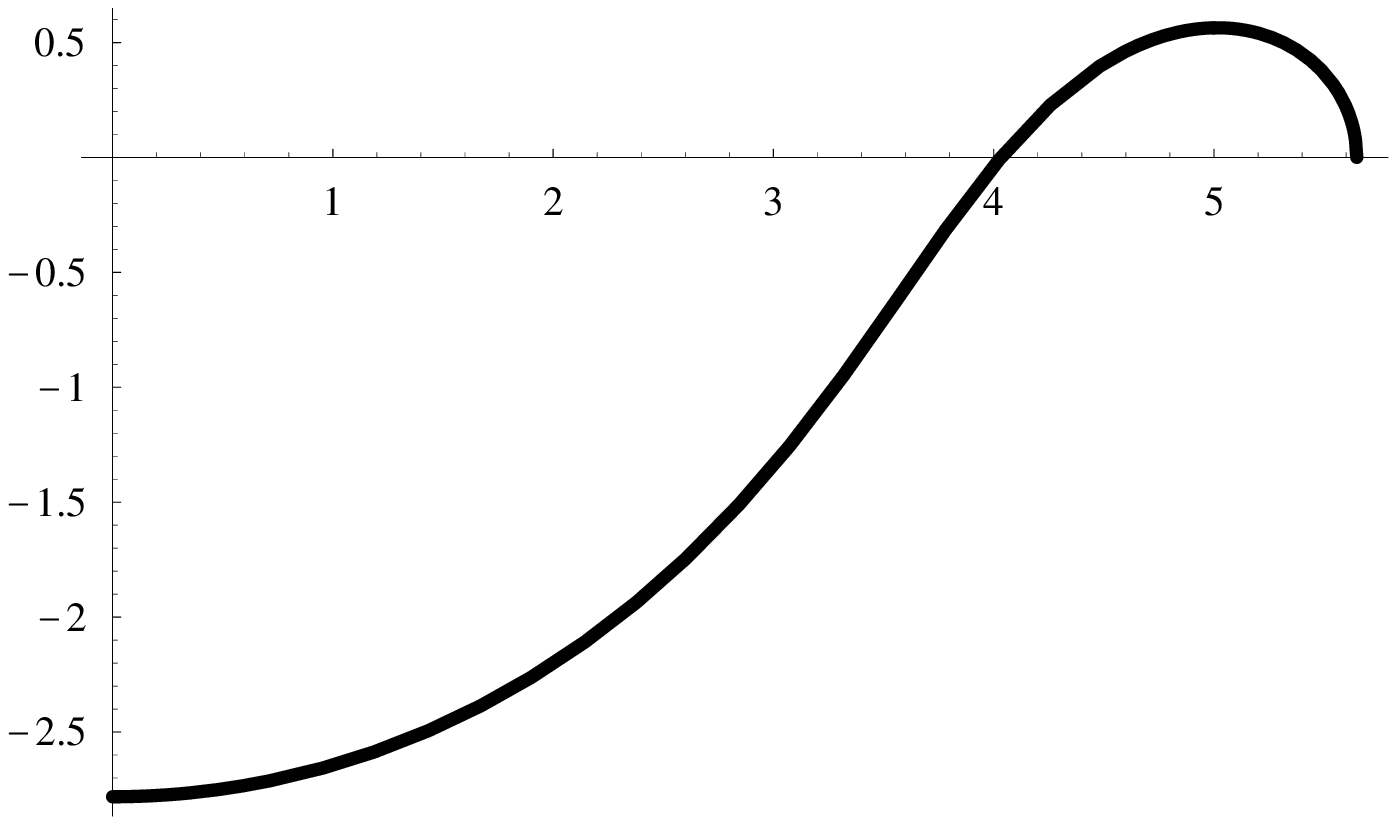)}
&\parbox{2cm}{\xpeinture 2cm by 1.4cm (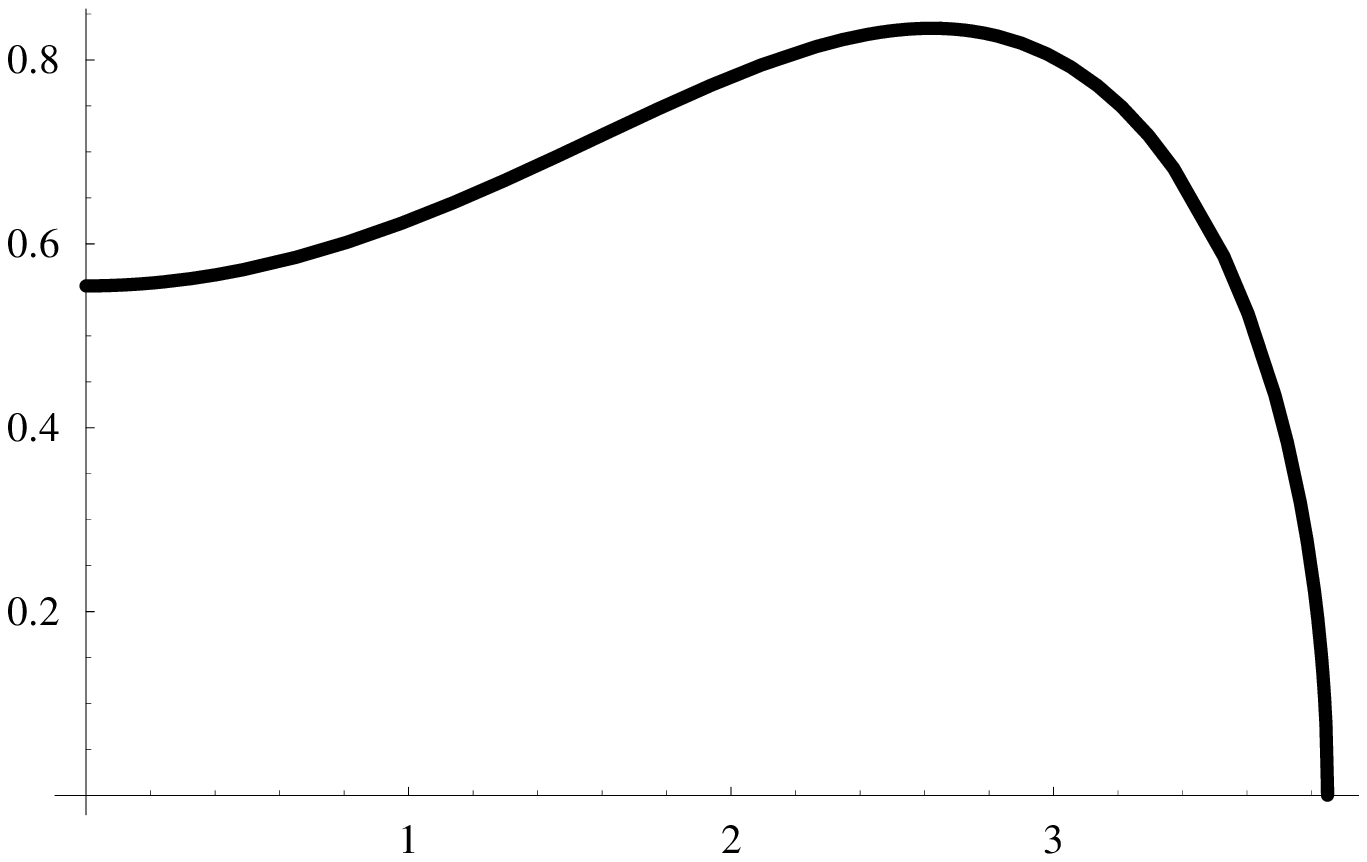)}
\\
\hline\hline
\end{tabular}

\smallskip

\noindent
\begin{tabular}{||ccc||}
\hline\hline
\multicolumn{3}{||c||}{\parbox{6.5cm}{
\begin{center}
Graph of $Q$ and positions of $w_0'$ \\
($c_0 = -1$, $\tlambda = -0.25$, $\tp = 1$) \\
\mbox{\xpeinture 6cm by 4cm (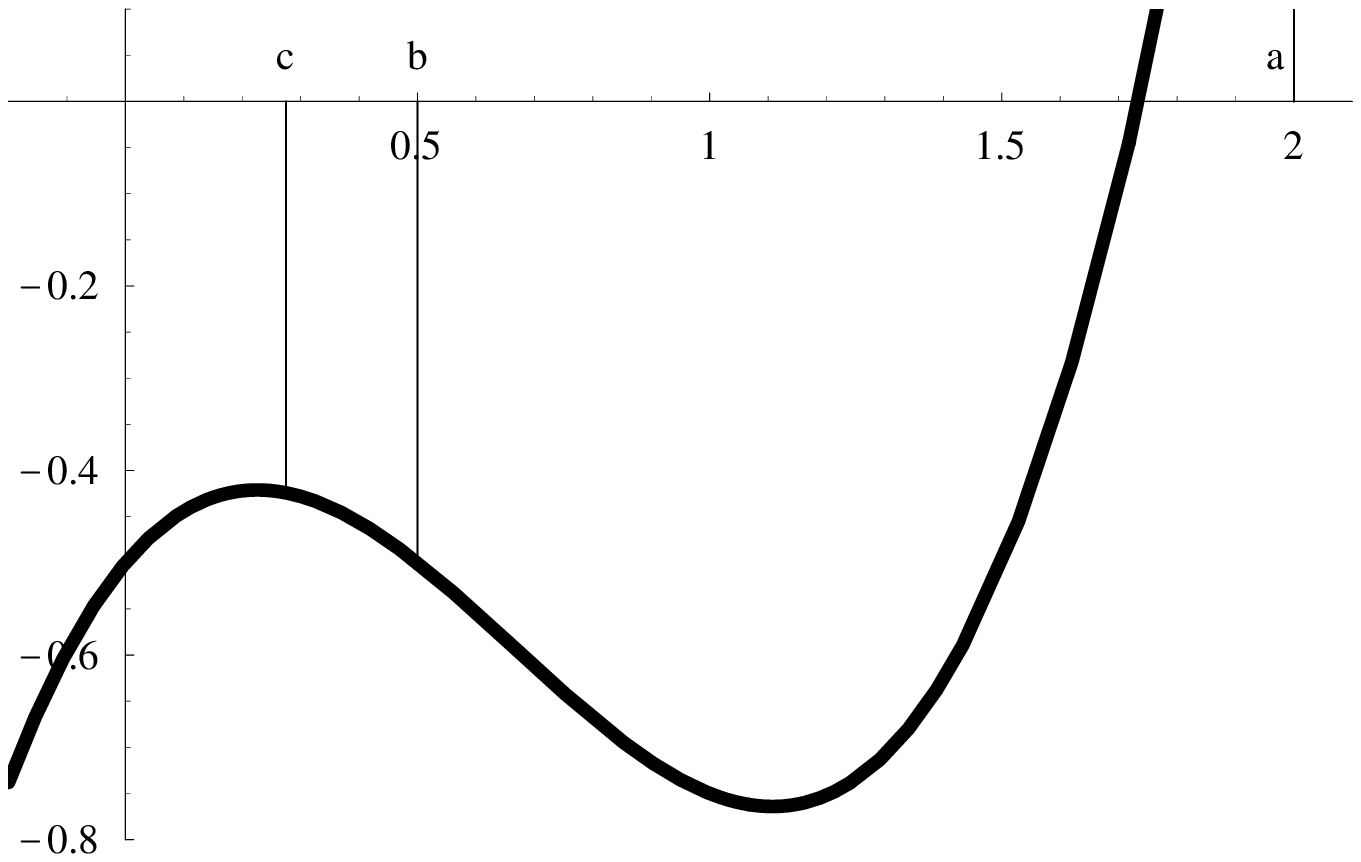)}
\end{center}
}} \\
\hline
\multicolumn{3}{||c||}{Graphs of solution $z$} \\
{\tiny (a) $w_0'=2$} 
& {\tiny (b) $w_0'=0.5$}
& {\tiny (c) $w_0'=0.275$}
\\
\parbox{2cm}{\xpeinture 2cm by 1.4cm (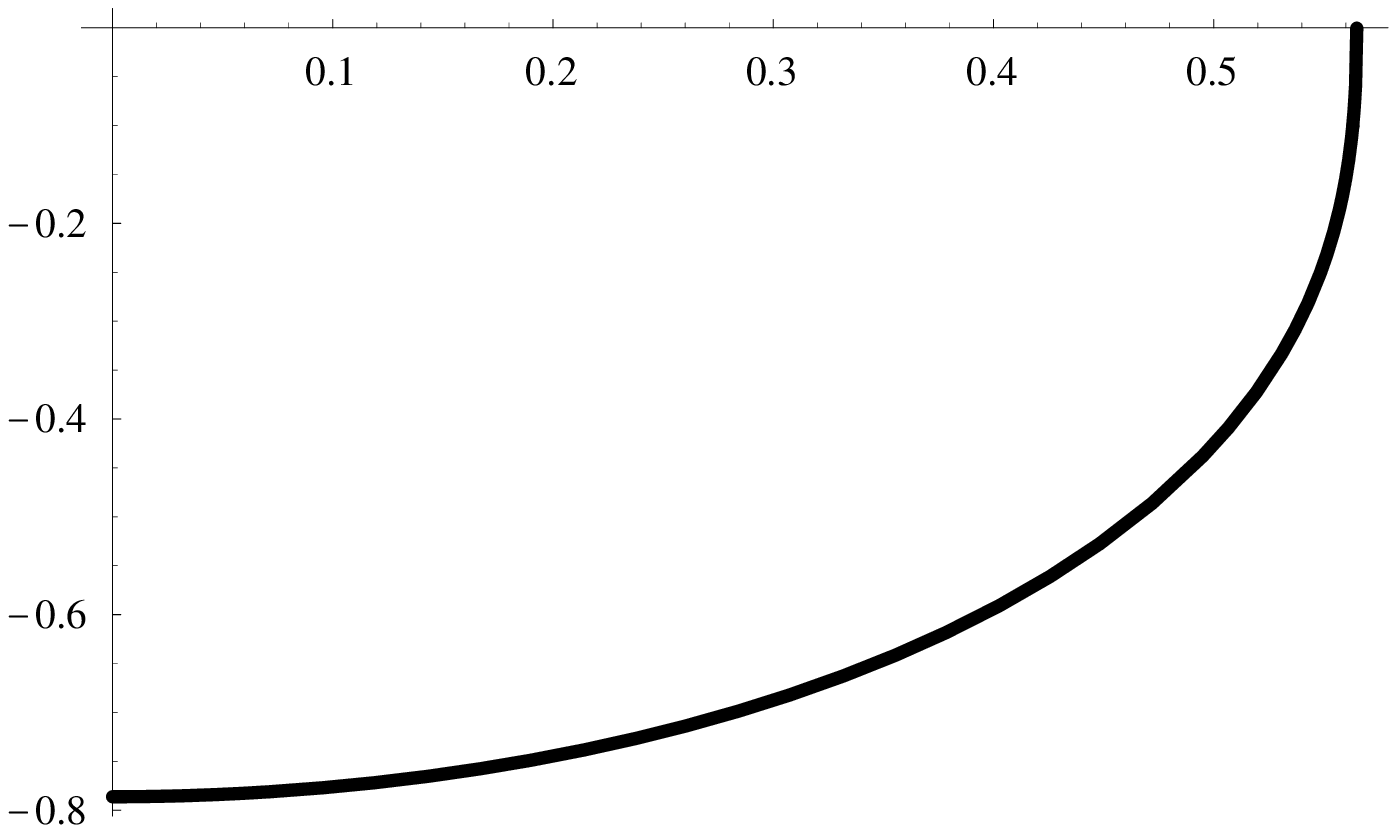)}
&\parbox{2cm}{\xpeinture 2cm by 1.4cm (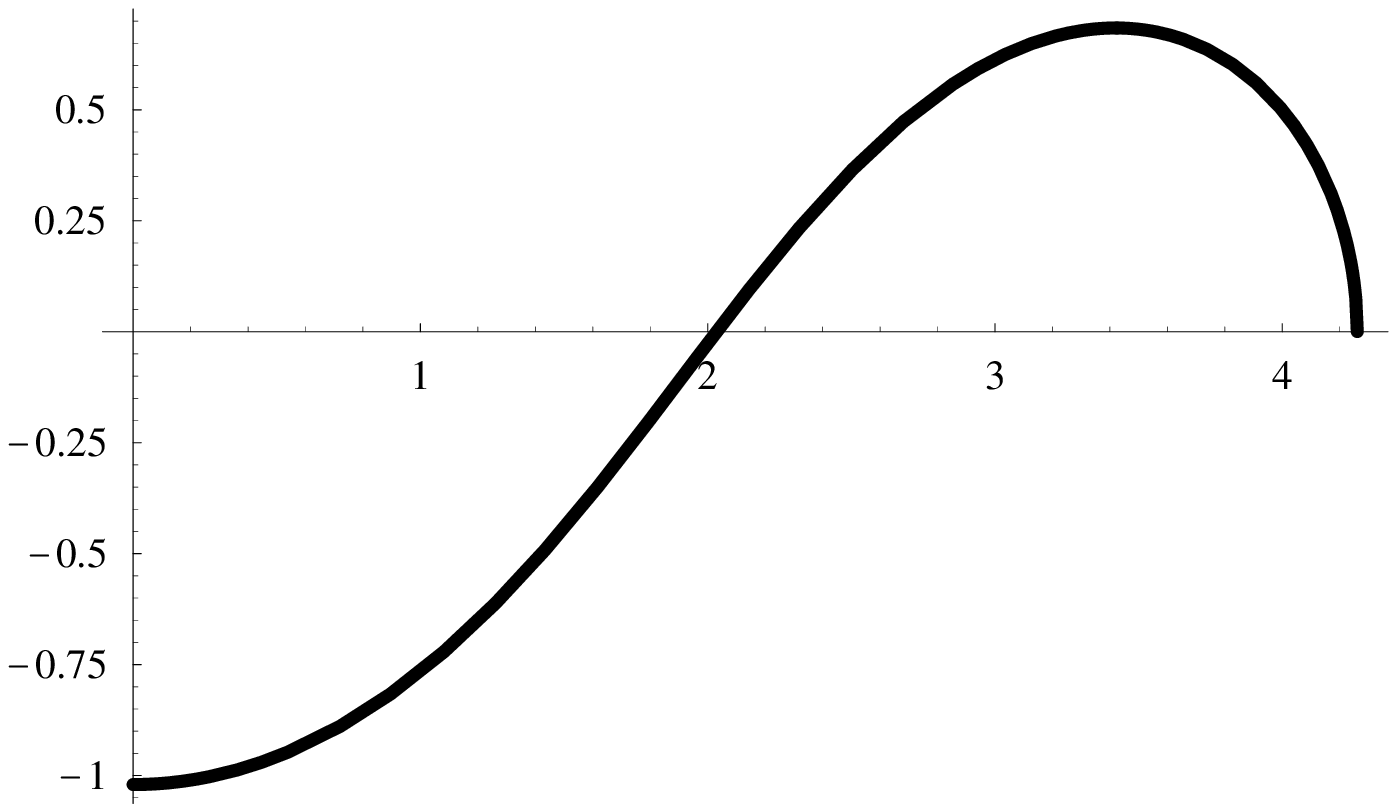)}
&\parbox{2cm}{\xpeinture 2cm by 1.4cm (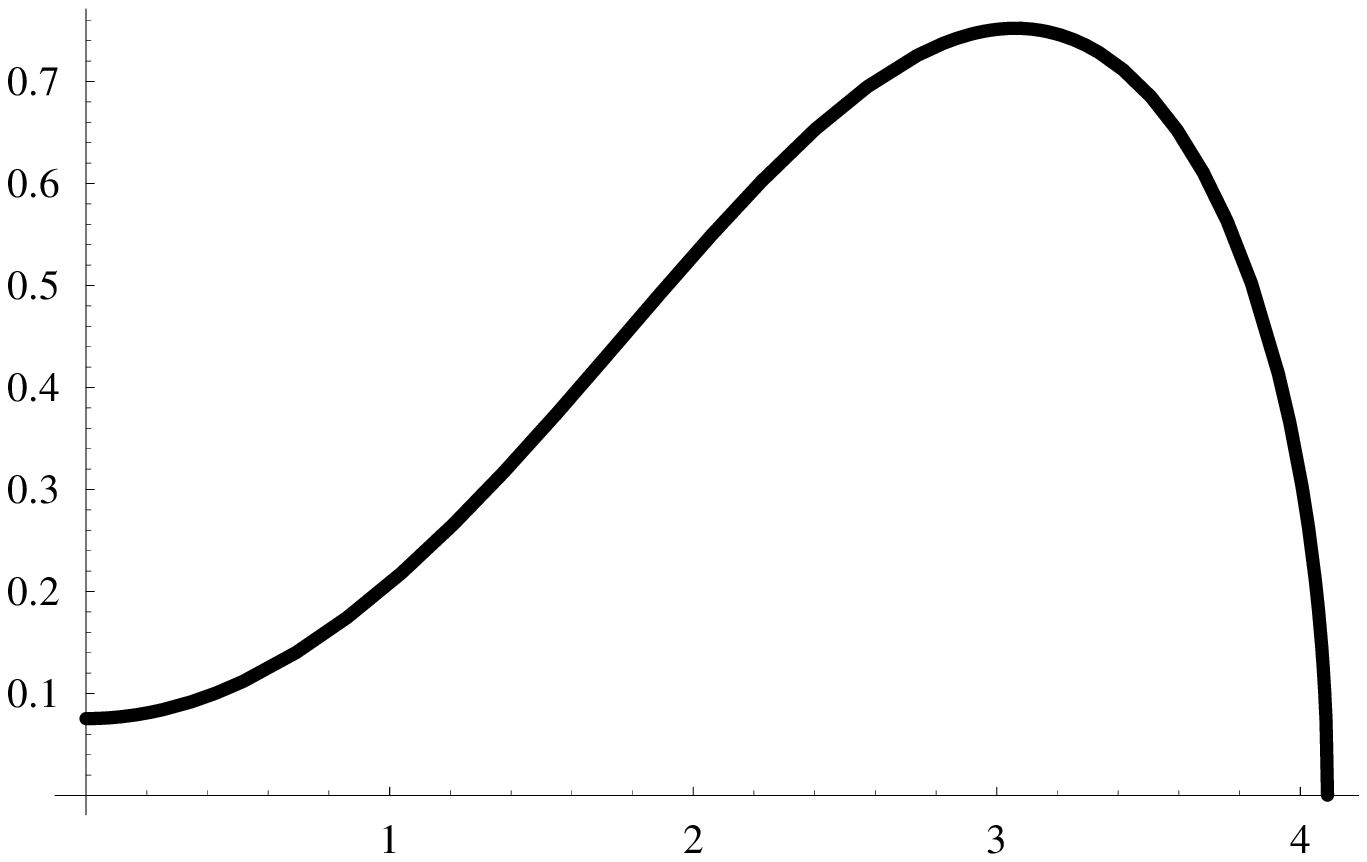)}
\\
\hline\hline
\end{tabular}

\setlength\baselineskip{10pt}

\end{document}